\documentclass{emulateapj}
\begin{document}

\def\etal{et al.\ }
\def\gtsima{$\; \buildrel > \over \sim \;$}
\def\ltsima{$\; \buildrel < \over \sim \;$}
\def\gsim{\lower.5ex\hbox{\gtsima}}
\def\lsim{\lower.5ex\hbox{\ltsima}}
\def\simgt{\lower.5ex\hbox{\gtsima}}
\def\simlt{\lower.5ex\hbox{\ltsima}}
\def\simpr{\lower.5ex\hbox{\prosima}}
\def\la{\lsim}
\def\ga{\gsim}
\def\CIV{C{\sc ~iv}  }
\def\SiIV{Si{\sc ~iv} }
\def\Lya{Ly$\alpha$~}
\def\msun{{M_\odot}}
\def\ie{{\frenchspacing\it i.e. }}
\def\eg{{\frenchspacing\it e.g. }}
\newcommand{\be}{\begin{equation}}
\newcommand{\ba}{\begin{eqnarray}}
\newcommand{\ee}{\end{equation}}
\newcommand{\ea}{\end{eqnarray}}
\def\msun{\,{\rm M_\odot}}

\def\Xtophe#1{\noindent{\bf[$\spadesuit$ #1]}}
\def\Bastien#1{\noindent{\bf[$\heartsuit$ #1]}}
\def\PPJ#1{\noindent{\bf[$\clubsuit$ #1]}}
\def\Evan#1{\noindent{\bf[$\diamondsuit$ #1]}}      

\title{The Clustering of Intergalactic Metals}

\author{
Christophe Pichon\altaffilmark{1,2,3},Evan Scannapieco\altaffilmark{4}, 
Bastien Aracil\altaffilmark{2}, Patrick Petitjean\altaffilmark{2,5},\\
Dominique Aubert\altaffilmark{1,2,3},
Jacqueline Bergeron\altaffilmark{2},
St\' ephane Colombi\altaffilmark{2,3}}
\altaffiltext{1}{Observatoire de Strasbourg, 11 rue de
l'Universit\' e, 67000 Strasbourg, France }
\altaffiltext{2}{Institut d'Astrophysique de Paris, 98 bis
boulevard d'Arago, 75014 Paris, France}
\altaffiltext{3}{Numerical Investigations in Cosmology
(N.I.C.), CNRS, France}
\altaffiltext{4}{Osservatorio Astrofisico di Arcetri, 
                     Largo E. Fermi 5, 50125 Firenze, Italy}
\altaffiltext{5}{LERMA, Observatoire de Paris, 61 avenue
de l'Observatoire, F-75014 Paris, France}

\begin{abstract}

We measure the spatial clustering of metals in the intergalactic
medium from $z=1.7$ to $3.0$, as traced by 643 \CIV and 104 \SiIV $N
\ge 10^{12}$ cm$^{-2}$ absorption systems in 19 high signal-to-noise
(40-80) and high resolution ($R=45000$) quasar spectra.  The number
densities and two-point correlation functions of both these species
are largely constant with redshift, suggesting the bulk of metal
ejection occurred at $z \ge 3.$ However, at $z \leq 1.9$ some
additional signature appears in the \CIV correlation function at 500
km/s, associated with four strong and peculiar systems.  At all
redshifts, the \CIV and \SiIV correlation functions exhibit a steep
rise at large separations and a flatter profile at small separations,
with an elbow occurring at $\sim 150$ km/s.  We show that these
properties are consistent with metals confined within bubbles with a
typical radius $R_s$ about sources of mass $\geq M_s$, and use
numerical simulations to derive best-fit values of $R_s \sim 2$
comoving Mpc and $M_s \sim 5 \times 10^{11} \msun$ at $z=3$.  This
does not exclude that metals could have been produced at higher
redshifts in smaller, but equally rare, objects.  At the level of
detection of this survey, IGM enrichment is likely to be incomplete
and inhomogeneous, with a filling factor $\sim 10\%.$

\end{abstract}


\section{Introduction}

No one knows where they came from, how they got there, or when it
happened, but quasar (QSO) absorption line studies have encountered
heavy elements in all regions of the tenuous intergalactic medium
(IGM) in which they were detectable (eg.\ Songalia \& Cowie 1996).
First, measurements of $N_{\rm CIV}/N_{\rm HI}$ indicated that
typically $[C/H] \simeq -2.5$ in somewhat overdense regions at $z
\simeq 3$ , with an order of magnitude scatter (eg.\ Rauch, Haehnelt,
\& Steinmetz 1997). Later, statistical methods pushed further, showing
that \CIV may be present in underdense regions (Ellison \etal 2000),
and that a minimum IGM metallicity of approximately $3\times 10^{-3}
Z_\odot$ is already in place at $z \approx 5$ (Songaila 2001;
hereafter S01).  Although the presence of metals in underdense regions
has not yet been firmly established (eg.\ Petitjean 2001; Carswell
\etal 2002; Schaye \etal 2003; Aracil \etal 2003a), their very
existence outside galaxies has profound cosmological implications.

As the long cooling times of large clouds of primordial composition
are shortened by even modest levels of enrichment (Sutherland \&
Dopita; Scannapieco \& Broadhurst 2001), IGM metals accelerated the
formation of massive ($\gsim 10^{12} \msun$) galaxies.  Similarly, the
violent events that ejected heavy elements from galaxies have
important implications for the thermal and velocity structure of the
IGM (eg.\ Gnedin \& Ostriker 1997), and would have exerted strong
feedback on dwarf galaxies forming nearby (eg.\ Thacker, Scannapieco,
\& Davis 2002).

Still, no one knows where they came from.  Although numerous
starburst-driven outflows have been observed at $3 \lesssim z \lesssim
5 $ (Pettini \etal 2001; Frye, Broadhurst, \& Benitez 2002), it is
unclear whether such objects are responsible for the majority of IGM
enrichment.  In fact a variety of theoretical arguments suggest that
these galaxies represent only the tail of a larger population of small
``pre-galactic'' starbursts that formed earlier (Madau, Ferrara, \&
Rees 2001; Scannapieco, Ferrara, \& Madau 2002).  On the other hand
active galactic nuclei are observed to host massive outflows (Weyman
1997), whose contribution remains unknown.  Finally, theoretical
studies have suggested that the primordial generation of metal-free
stars were very massive (eg. Schneider \etal 2002), resulting in a
profusion of tremendously powerful pair-production supernovae that
distributed metals at $z \gsim 15$ (Bromm \etal 2003).

This paper describes the first direct attempt to use clustering
measurements to discriminate between these disparate models.  Making
use of 643 \CIV and 104 \SiIV absorption systems, measured in 19 high
signal-to-noise quasar spectra, we are able to place strong
constraints on the spatial distribution of intergalactic metals at
intermediate redshifts ($1.5 \leq z \leq 3.1$ and a detection limit of
$N_{CIV} \geq 10^{12}$ cm$^{-2}$).  Whatever the objects that enriched
the IGM, it is clear that they formed in the densest regions of space,
regions that were highly clustered.  As this ``geometrical biasing''
is a systematic function of mass (eg.\ Kaiser 1984), the large-scale
clustering of metal-line systems encodes valuable information on the
masses of the sources from which they were ejected. Likewise, 
we are able to relate the small-scale clustering of metals to
the typical radius enriched by each source.

The structure of this work is as follows.  In \S2 we summarize the
properties of our data set and reduction methods.  In \S3 we present
the number density of \CIV and \SiIV absorption systems and their two
point correlations.  In \S4 we compare these correlations with
simulations to constrain the masses and ejection energies of the
sources responsible for IGM metal enrichment.  A discussion is given
in \S5. 
 
\section{Data Set and Reduction}

\subsection{Observations}

The ESO Large Programme ``The Cosmic Evolution of the IGM'' was
devised to provide a homogeneous sample of QSO sight-lines suitable
for studying the IGM from $z$ = 1.7$-$4.5.  High resolution
($R$~$\sim$~45000), high signal-to-noise (40 and 80 per pixel at 3500
and 6000~\AA, respectively) spectra were taken over the wavelength
ranges 3100--5400 and 5450--9000~\AA, using the UVES spectrograph on
the Very Large Telescope.  Although the complete emission redshift
range is covered, emphasis is given to lower redshifts, and the \CIV
and \SiIV metal lines discussed in this paper were well-detected over
the redshift ranges of 1.5-3.0 and 1.8-3.0 respectively, as described
in Table 1.

Observations were performed in service mode over two years.  The data
were reduced using the UVES context of the ESO MIDAS data reduction
package, applying the optimal extraction method, and following the
pipeline reduction step by step.  The details of this reduction and
spectrum normalization are described in further detail in Aracil \etal
(2003a).

\vspace{.2in}
\begin{scriptsize}
\begin{minipage}{\hsize}
\centerline{
\begin{tabular}{llcc}
\hline
\qquad Name & $z_{\rm em}$ & \multicolumn{2}{c}{Coverage}   \\
     &              &   C~{\sc iv} & Si~{\sc iv}  \\
\hline\\
PKS2126$-$158  & 3.280 & 2.36$-$3.24 & 2.74$-$3.24\\
Q0420$-$388    & 3.117 & 2.23$-$3.08 & 2.59$-$3.08 \\
HE0940$-$1050  & 3.084 & 2.21$-$3.04 & 2.56$-$3.04\\
HE2347$-$4342  & 2.871 & 2.04$-$2.83 & 2.38$-$2.83\\
HE0151$-$4326  & 2.789 & 1.97$-$2.75 & 2.31$-$2.75\\
Q0002$-$422    & 2.767 & 1.96$-$2.73 & 2.29$-$2.73\\
PKS0329$-$255  & 2.703 & 1.91$-$2.66 & 2.23$-$2.66\\
Q0453$-$423    & 2.658 & 1.87$-$2.62 & 2.19$-$2.62\\
HE1347$-$2457  & 2.611 & 1.83$-$2.57 & 2.15$-$2.57\\
HE1158$-$1843  & 2.449 & 1.71$-$2.41 & 2.01$-$2.41\\
Q0329$-$385    & 2.435 & 1.70$-$2.40 & 2.00$-$2.40\\
HE2217$-$2818  & 2.414 & 1.68$-$2.37 & 1.98$-$2.37\\
Q1122$-$1328   & 2.410 & 1.68$-$2.37 & 1.98$-$2.37\\      
Q0109$-$3518   & 2.404 & 1.67$-$2.36 & 1.97$-$2.36\\
HE0001$-$2340  & 2.263 & 1.56$-$2.22 & 1.84$-$2.22\\
PKS0237$-$23   & 2.222 & 1.53$-$2.18 & 1.81$-$2.18\\
PKS1448$-$232  & 2.220 & 1.53$-$2.18 & 1.81$-$2.18\\
Q0122$-$380    & 2.190 & 1.50$-$2.15 & 1.78$-$2.15\\
HE1341$-$1020  & 2.135 & 1.46$-$2.10 & 1.74$-$2.10\\
\hline\\
\end{tabular}
}
\end{minipage}
\end{scriptsize}
{\small \qquad {\sc Table}~1.  List of lines of sight.  Here $z_{\rm em}$
is the quasar emission redshift. Only \CIV and \SiIV systems detected
redward of the \Lya forest and at least 3000 km/s blueward of
$z_{\rm em}$ are used.}

\subsection{Line Identification}

Metal-line systems were identified using an automated two-step
procedure.  For each species we applied a pixel-by-pixel procedure
which compared the spectrum with one that was rescaled and shifted
according to the ratio of the oscillator strengths, extracting the
minimal flux and accounting for blends.  A detection threshold was
then applied to these spectra, such that only absorption features with
equivalent widths (EWs) larger than 5 times the noise rms were
accepted.  Finally a set of physical criteria were applied, to
eliminate false detections.

For each identified system the final decomposition into subcomponents
was carried out using VPFIT (Carswell \etal 1987).  For each
component, initial $z$ values were taken from the central wavelengths
of the sub-features, while initial values for the Doppler parameter
$b$ and column density $N$ were computed directly from the EW and peak
absorption of each sub-feature.  This procedure is described in detail
in Aracil \etal (2003b) and has been tested on simulated spectra,
doing well for all systems with realistic values of $N$ and $b$.

Finally we applied a set of four cuts to the automated list generated
by VPFIT: $\log N \, [{\rm cm}^{-2}] \, \geq 12$ due to the detection limit
of our procedure, $b \geq 3 $ km/s to avoid false detections due to
noise spikes, $\log N [{\rm cm}^{-2}] \leq 16$ to remove two very badly
saturated subcomponents, and $b \leq 45$ km/s to avoid false systems
due to errors in continuum fitting.  Our cuts resulted in a final data
set of 643 \CIV and 104 \SiIV systems, drawn from an original sample
of 748 and 149 respectively.

\section{Analysis}

\subsection{Distribution Functions}

This  sample was  first  used  to compute  the  column density  distribution
function $f(N)$.  Following  Tytler (1987), $f(N)$ is defined  as the number
of  absorbing systems per  unit column  density and  per unit  redshift path
$X(z)  \equiv \frac{2}{3}\left[(1+z)^{3/2}-1  \right]$.  In  an $\Omega_m=1$
universe,  which  is a  reasonable  approximation  over  the redshift  range
considered, $f(N)$ does not evolve  for a population whose physical size and
comoving space density are constant.

{\epsscale{1.05} \plotone{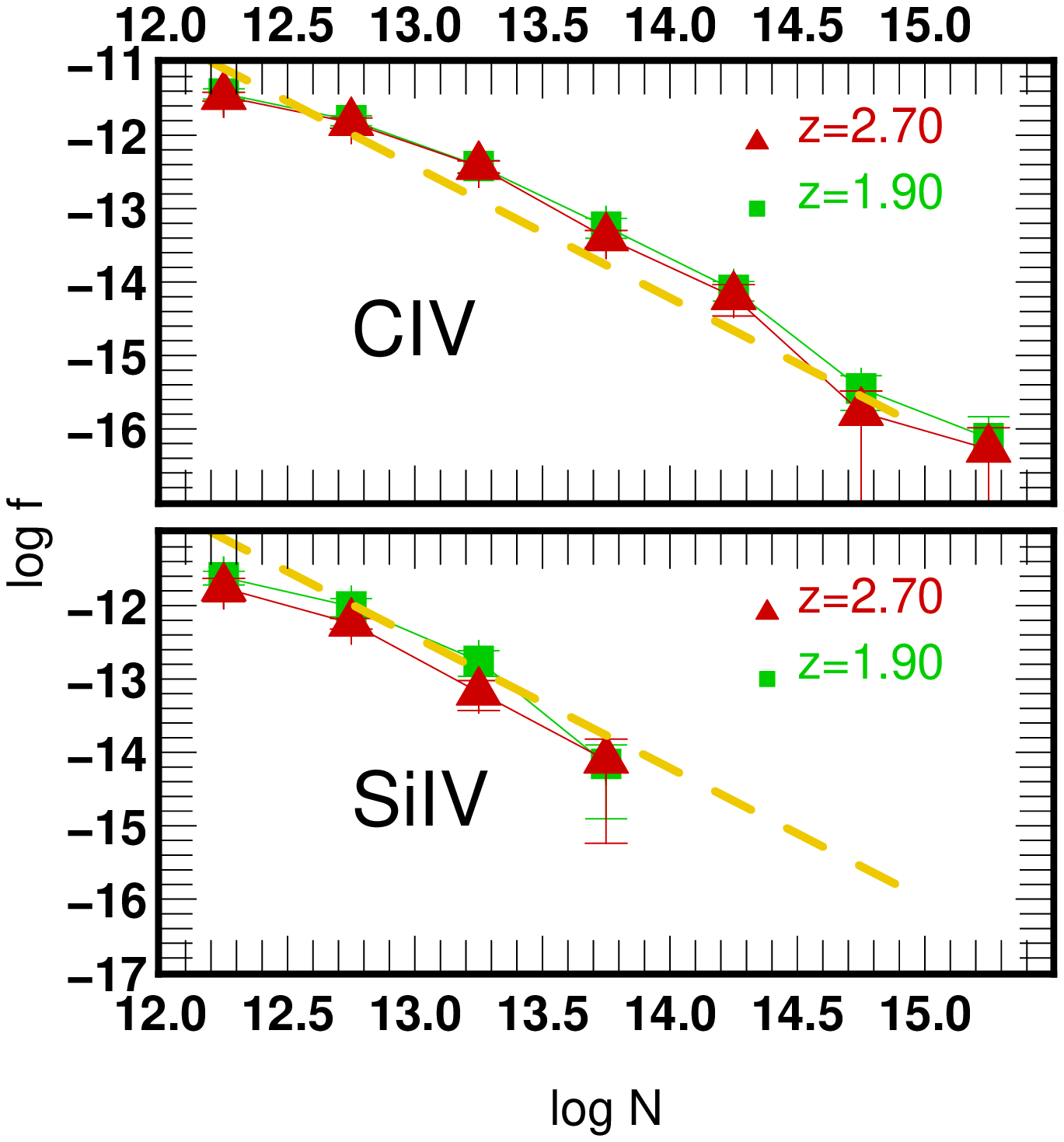}} {\small {\sc
Fig.}~1.---Column density distribution of \CIV (upper panel) and \SiIV
(lower panel) absorption systems.  In each panel, systems are divided
into two redshift bins: $1.5 \leq z \leq 2.3$ (squares) and $2.3 \leq
z \leq 3.1$ (triangles).  Column density bins are $10^{0.5} N$
cm$^{-2}$ wide and $\pm 1 \sigma$ error bars are given, computed from
the variance of the data between spectra.  The dashed lines are the
power law fit measured in S01.\\}

In Fig.\ 1 we plot $f(N)$ for both \CIV and \SiIV systems, dividing
our sample into two redshift bins (1.5-2.3 and 2.3-3.1). Both species
are consistent with a lack of redshift evolution, as found by previous
studies of \CIV and \SiIV absorption systems (S01; Ellison \etal
2000).  Furthermore the overall density distribution of \CIV is
consistent with a power-law of the form $f(N) \propto N^{-1.8}$ as fit
by S01.  From this figure, we see that some incompleteness sets in
around $10^{12.5}$ cm$^{-2}$.  While fewer in total, \SiIV systems are
also consistent with a lack of evolution, and follow a similar power
law with a lower overall magnitude. 

\subsection{Clustering of Metal Absorption Systems}

Having constructed a sample of well-identified metal absorption
systems, whose distributions are consistent with previous
measurements, we then computed their two-point correlation function in
redshift space, $\xi(v)$.  This quantity was last studied by Rauch
\etal (1996) and Boksenberg, Sargent, \& Rauch (2003) who noted a
marked similarity between $\xi(v)$ of \CIV and Mg~II (see also
Petitjean \& Bergeron 1990, 1994).  For each quasar, we computed a
histogram of all velocity separations and divided by the number
expected for a random distribution.  We then combined these
distributions, weighting them according to the number of pairs in each
quasar, yielding $\xi(v)+1$.  The corresponding error bars were
computed according to the appropriate weighted variance between
quasars.

{\epsscale{1.05} \plotone{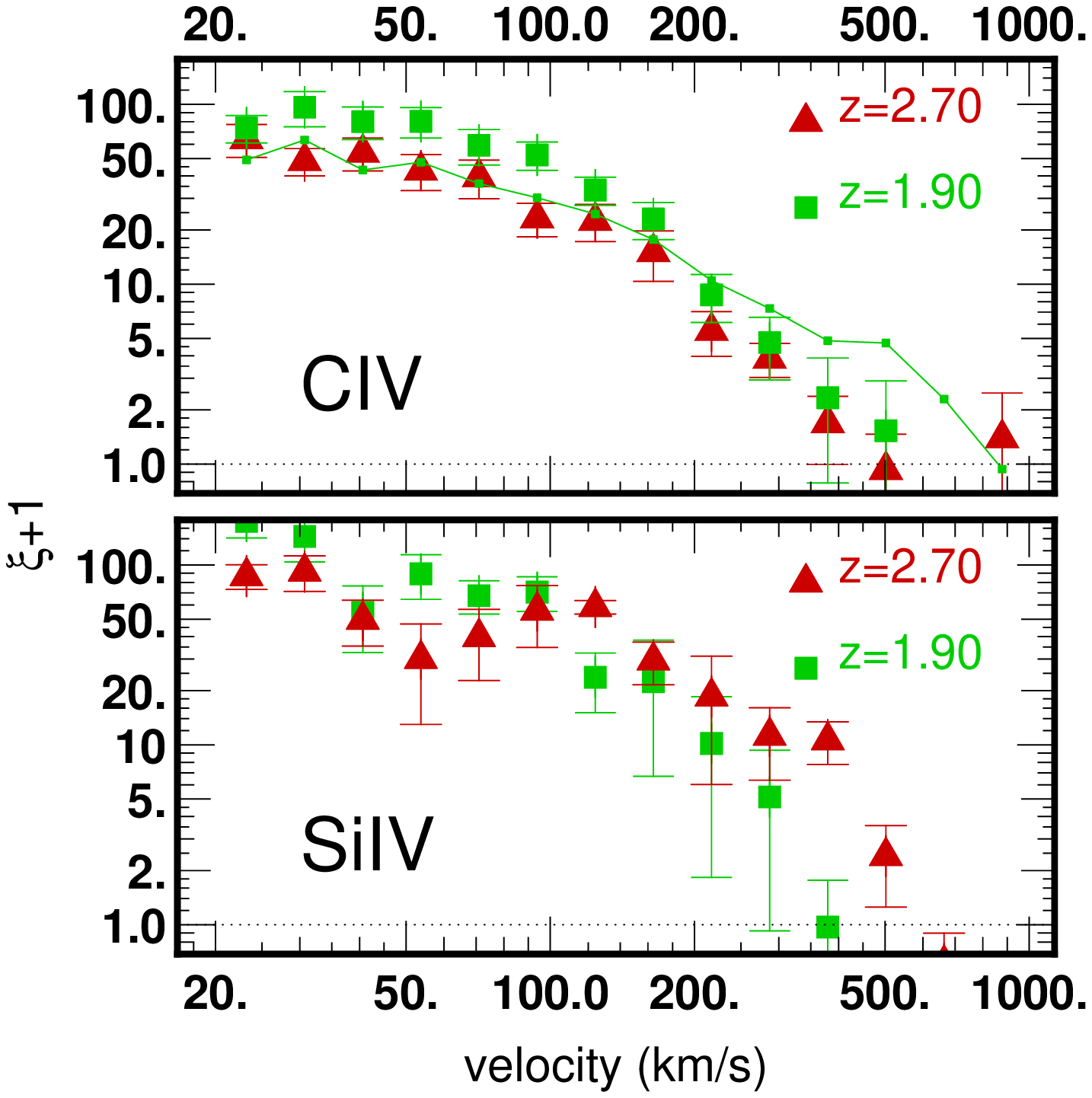}}  {\small {\sc
Fig.}~2.---Two point correlation function of \CIV (upper panel) and
\SiIV (lower panel) absorption systems.  In each panel our systems
have been divided into two redshifts bins, with symbols as in Fig.\ 1.
The solid line in the upper panel includes the four unusual
low-redshift systems (see text). \\}

The resulting correlation functions are shown in Fig.\ 2, again split
into two redshift bins.  While the spatial distribution of both
species shows no evolution at separations $\leq 400$ km/s, our
comparison uncovers a low-redshift feature in the 500 km/s \CIV bin.
This we have traced back to arising from four extended systems amongst
our 19 QSO sightlines, and seems to be due to some peculiarity in the
internal structure of these systems.  These will be studied in detail
in a future publication, and are excluded from our model comparisons
below, although this cut has no effect on our final fits.

The overall shape and amplitude of the \CIV and \SiIV correlation
functions are strikingly similar and are consistent to within the
\SiIV measurement errors.  Both functions exhibit a steep rise at
large separations and a flatter profile at small separations, with an
elbow occurring at $\sim 150$ km/s. Unlike \Lya absorption systems
(see Cristiani \etal 1997), the correlation of \CIV does not seem to
depend strongly on absorption column densities.  While choosing subsets
below a fixed $N$ increases the noise of our measurements, the shape
of $\xi(v)$ is preserved to within error bars.
\vskip 0.5cm

\section{Constraints on Metal Sources}

The two-slope shape of the measured correlation functions is highly
suggestive of a picture in which metal bubbles of a typical size
(associated with the small scale clustering) are generated about
objects of a typical mass (whose geometrical bias is associated with
the large-scale clustering) (see also Petitjean \& Bergeron 1990).  In
order to explore this connection further, we generated a simple model,
based on a dark-matter-only simulation.  As no evolution was detected,
and the properties of both species were similar, we focused our
attention on comparing the properties of \CIV absorbers with simulated
groups at a single representative redshift of three.

We adopted a Cold Dark  Matter cosmological model with $h=0.7$, $\Omega_0$ =
0.3,  $\Omega_\Lambda$ =  0.7, $\Omega_b  =  0.05$, $\sigma_8  = 0.87$,  and
$n=1$,  where $\Omega_0$,  $\Omega_\Lambda$,  and $\Omega_b$  are the  total
matter,  vacuum, and  baryonic densities  in units  of the  critical density
($\rho_c$), $\sigma_8^2$  is the variance  of linear fluctuations on  the $8
h^{-1}{\rm  Mpc}$ scale, and  $n$ is  the ``tilt''  of the  primordial power
spectrum.  We made  use of a simulation with  $512^3$ dark matter particles,
contained  with a  box $50/h$  comoving Mpc  on a  side.  The  mass  of each
particle  was  $7.7  \times  10^{7}  \msun$,  which  gives  a  minimum  mass
resolution  for our  group finding  of $10^{9.5}  \msun$ as  we  select only
groups  with 40  or more  particles.  The  initial conditions  were produced
using GRAFIC (Bertschinger 2001) and were used as the inputs to the parallel
version of GADGET (Springel \etal 2001).  Halo detection was performed using
the  HOP algorithm  (Eisenstein  \&  Hut 1998)  with  the cuts  $\delta_{\rm
peak}=160$, $\delta_{\rm saddle}=140$, and $\delta_{\rm outer}=80$.

The resulting list of groups was then used to construct a simple
simulated picture of IGM metal enrichment.  We selected all $z=3$
groups above a threshold group mass $M_s$ as markers of the centers of
enrichment events, about which we painted spheres of radius $R_s$.
Note that our choice of this set is not meant to imply that enrichment
occurred at $z=3$, but rather that it occurred at an unknown redshift
higher than the observed range, centered on groups whose large-scale
clustering was equivalent to $z=3$ objects of mass $M_s$.  All smaller
groups within such spheres were associated with \CIV absorbers, each
with an overall impact parameter equal to a factor $\tilde b$ times
its dark matter radius.  The observed number of \CIV systems per unit
redshift path was well-reproduced by fixing $\tilde b = 10.$

\setcounter{figure}{2} 

\begin{figure*}[t]
\vspace{60mm}
\includegraphics{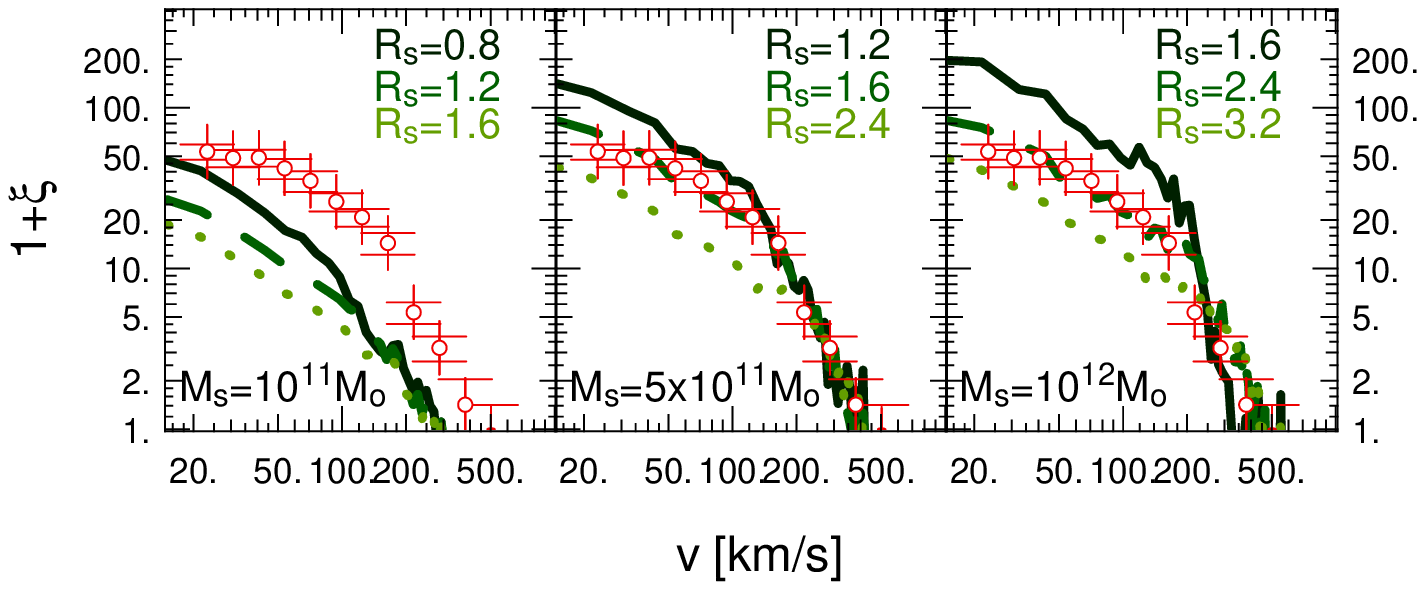}
\caption{Comparison between detected \CIV correlation functions
(points) and simulated correlations (lines).  In each panel $M_s$ is fixed,
$R_s$ is varied over three representative values, and $\tilde b = 10$.}
\end{figure*}

Thirty-two planes  were taken, and  for each plane all  512 sightlines
were drawn;  the average two-point correlation was  then compared with
detected $\xi$  values, as shown in  Fig.\ 3.  As  expected, the large
scale correlations are almost  completely dependent on $M_s$ while the
position of the  elbow in $\xi(v)$ is set  by $R_s$.  Our measurements
are  clearly suggestive  of $R_{s}=1.6$  Mpc and  $M=5  \times 10^{11}
M_{\odot}$ $z=3$ objects, corresponding to a filling factor of $f \sim
10 \%$, at  the level of detection of our  survey.  Varying $\tilde b$
over a wide range of values has only a minor impact on this comparison,
 changing the  inner slope of $\xi(v)$ slightly  while giving a
much poorer fit for the overall number of systems.

A straightforward geometric argument allows us to associate $R_{s}$
with a wind ejection energy per object, $E_s$.  Here we adopt a simple
picture in which the average bubble expands according to a
Sedov-Taylor solution (Sedov 1959).  In this case $E_s$ is related to
the bubble radius as $E_s = {48 \pi R^5 \rho t^{-2}}/125$, where $t$
is the expansion time and $R$ is in physical units such that $R =
R_s/(1+z_o)$, where is the redshifts of observation, respectively.  If
we assume for simplicity that $t$ is approximately the Hubble time
(2.3 Gyrs at $z=3$ in our cosmological model) and that the shell
expands into material at the mean density at this redshift, this gives
$E = 1.6 \times 10^{58}$ erg, which is equal to the kinetic energy
input from roughly $1.3 \times 10^7$ TypeII supernovae (SNIIe).  This
is roughly consistent with the overall number of \CIV systems, which
can be converted into a cosmological density of $\Omega_{\CIV} \approx
5 \times 10^{-8}$ as described in S01.

\section{Discussion}

We have studied the clustering of 643 \CIV and 104 \SiIV systems
detected in UVES observations of 19 high-$z$ QSOs using an automatic
procedure.  The independently derived correlation functions of \CIV
and \SiIV are similar, giving confidence in our automated approach.
The shape of the correlation function is consistent with a picture
where metals are confined within bubbles of $R_s \sim 2$ Mpc about
$M_s \sim 5 \times 10^{11} \msun$ halos at $z=3$.  This implies that
the filling factor of the metals is only 10\% at the detection limit
of the survey.

As our overall mass constraint is derived only from the geometrical
bias of the sources, however, we cannot exclude models in which
equally rare, but somewhat smaller higher-redshift objects are
responsible for metal enrichment.  Furthermore, the observed number
density of \CIV shows little evolution at $z \leq 3$ as at higher
redshift (eg.\ S01).  While more detailed investigations are
necessary, it remains that all such models will be strongly
constrained by the clustering of intergalactic metals.

\acknowledgements

We are grateful to J.~Heyvaerts, C.~Mallouris D.~Pogosyan,
E.~Rollinde, R.~Teyssier, \& E.~Thi\'ebaut.  ES was supported in part
by an NSF MPS-DRF fellowship.  This work is based on observations
collected through ESO project ID No.  166.A-0106.  The computational
resources were made available to us by CINES.

\end{document}